\documentclass[reprint]{revtex4-1}
\usepackage{amssymb}

\usepackage{dcolumn}
\usepackage{amsmath}
\usepackage{color}
\usepackage{ulem}

\usepackage{graphicx}
\usepackage{epstopdf}

\usepackage{bm}
\usepackage{amssymb}
\usepackage{soul}
\usepackage{color}

\def\be{\begin{equation}}
\def\ee{\end{equation}}
\def\bea{\begin{eqnarray}} 
\def\eea{\end{eqnarray}}

\begin{document}


\title{Van der Waals Schottky barriers as interface probes of the correlation between chemical potential shifts and charge density wave formation in 1T-TiSe$_2$ and 2H-NbSe$_2$}


\author{Ang J. Li$^{1\dagger}$, Xiaochen Zhu$^{1\dagger}$, Daniel Rhodes$^{2}$, Christopher C. Samouce$^{1}$, Luis Balicas$^{2}$ and Arthur F. Hebard$^{1}$}
\affiliation{$^{1}$Department of Physics, University of Florida, Gainesville, FL 32611, USA\\
$^{2}$National High Magnetic Field Laboratory, Florida State University, Tallahassee, FL 32310, USA}

\date{\today}

\begin{abstract}
Layered transition metal dichalcogenide (TMD) materials, i.e. 1T-TiSe$_2$ and 2H-NbSe$_2$, harbor a second order charge density wave (CDW) transition where phonons play a key role for the periodic modulations of conduction electron densities and associated lattice distortions. We systematically study the transport and capacitance characteristics over a wide temperature range of Schottky barriers formed by intimately contacting freshly exfoliated flakes of 1T-TiSe$_2$ and 2H-NbSe$_2$ to \textit{n}-type GaAs semiconductor substrates. The extracted temperature-dependent parameters (zero-bias barrier height, ideality and built-in potential) reflect changes at the TMD/GaAs interface induced by CDW formation for both TMD materials. The measured built-in potential reveals chemical-potential shifts relating to CDW formation. With decreasing temperature a peak in the chemical-potential shifts during CDW evolution indicates a competition between electron energy re-distributions and a combination of lattice strain energies and Coulomb interactions.  These modulations of chemical potential in CDW systems, such as 1T-TiSe$_2$ and 2H-NbSe$_2$ harboring second-order phase transitions, reflect a corresponding conversion from short to long-range order. 
\end{abstract}

\maketitle
\section{Introduction}
Understanding the mechanism of charge density wave (CDW) formation in low-dimensional correlated systems is of fundamental importance in condensed matter science, where studies of layered transition metal dichalcogenides (TMDs) harboring CDWs, such as 1T-TiSe$_2$ and 2H-NbSe$_2$\cite{tise1, tise2, nbse1, nbse2}, have revealed critical insights. The conventional weak-coupling mechanism driving the CDW state, known as the Peierls instability in 1-D, derives from Fermi surface nesting\cite{peierls} in which the energy gain associated with the formation of an energy gap separating filled and empty states is greater than the energy cost of forming a lattice distortion.\cite{gruner} On the other hand CDW formation in 1T-TiSe$_2$ and 2H-NbSe$_2$ is believed to be a consequence of strong-coupling\cite{tise strong1, bond tise, nbse2, nbse strong1, nbse strong2, nbse strong3}, where electron-electron or electron-phonon coupling is considered to be essential.

Direct observations by scanning tunneling spectroscopy/microscopy (STS/STM), angle-resolved photoemission spectroscopy (ARPES), X-ray powder diffraction (XRD), etc, have facilitated a better understanding of the physics of strong-coupling CDW systems. For example, in 2H-NbSe$_2$ with a temperature-independent CDW wave vector, a short range CDW sets in at a temperature almost three times higher than the CDW transition temperature $T_{\rm{CDW}}$ (33\thinspace K) where long range CDW order dominates\cite{nbse1}. Accordingly, for 2H-NbSe$_2$ the amplitude of the CDW order parameter is well defined within nanosized domains at higher temperature\cite{nbse1} and the phase coherence gradually increases as the temperature is lowered towards an onset of long-range order with global phase coherence at $T_{\rm{CDW}}$\cite{nbse2}. The TMD, 1T-TiSe$_2$, exhibits a different behavior. Here, although a short range coherent CDW state has been reported as well\cite{tise short1, tise short2}, the mechanism for CDW formation is associated with the opening of an indirect gap accompanied by the formation of excitons\cite{tise condensate}. The amplitude of the order parameter is proportional to the gap and thus with decreasing temperature (for $T < T_{\rm{CDW}}$) the amplitude increases towards saturation. For $T > T_{\rm{CDW}}$, 1T-TiSe$_2$ is a semimetal and the CDW does not exist.  


Transport measurements on CDW materials have traditionally provided evidence for a translational motion of the bulk CDW condensate in the presence of pinning to the underlying lattice\cite{gruner}.  In this work we report systematic temperature-dependent current-voltage ($I$-$V$) and capacitance-voltage ($C$-$V$) measurements on Van der Waals junctions of 1T-TiSe$_2$ and 2H-NbSe$_2$ intimately contacted to moderately-doped \textit{n}-type GaAs substrates. Specifically we find that mechanically exfoliated thin flakes of 1T-TiSe$_2$ and 2H-NbSe$_2$ form high quality Schottky barriers when placed into intimate contact with \textit{n}-GaAs substrates with a nominal room-temperature doping concentration of 4.1$\times$ 10$^{16}$ cm$^{-3}$. Advantageously our Schottky junctions serve as surface state probes that are sensitive to quasistatic chemical potential shifts that are associated with CDW formation in bulk. 
\section{Sample fabrication and experimental methods}

\begin{figure}
\hspace*{-0.15in}
\includegraphics[width=0.5\textwidth]{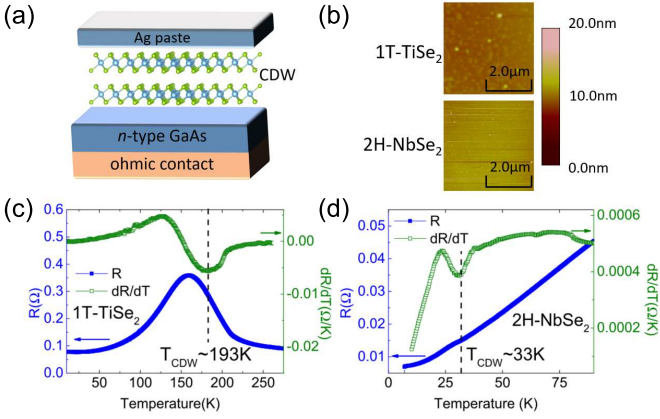}
\caption{(a): Schematic of sample structure with top and back contact. (b): Atomic force microscope (AFM) images of the cleaved surfaces for thin flakes of 1T-TiSe$_2$ and 2H-NbSe$_2$. (c)--(d): In-plane resistance and temperature derivative of the resistance of thin flakes of 1T-TiSe$_2$ and 2H-NbSe$_2$ plotted as a function of temperature. Black dashed vertical lines indicate the $T_{\rm{CDW}}$ for 1T-TiSe$_2$ and 2H-NbSe$_2$, respectively.}
\label{supfigure1}
\end{figure}

Mechanically exfoliated thin flakes of 1T-TiSe$_2$ and 2H-NbSe$_2$, with nominal thicknesses in the range 5-20\thinspace $\rm{\mu}$m, were transferred onto chemically cleaned \textit{n}-type GaAs wafers as shown in Fig.~\ref{supfigure1}(a). An RCA-I clean followed by a 3:1:50 HNO$_3$:HF:H$_2$O rinse for 2 minutes removed most native organics and oxides on the GaAs. According to our Hall measurements, the commercially available Si-doped GaAs(100) wafer had a nominal doping concentration of 4.1$\times$ 10$^{16}$ cm$^{-3}$ at room temperature. Low resistance ohmic contacts to the GaAs wafers, robust to temperatures as low as 5\thinspace K, were made by rapid thermal annealing using a series of previously described recipes\cite{Ang}. As shown in Figs.~\ref{supfigure1}(b) and (c), clean and flat flakes of cleaved 1T-TiSe$_2$ and 2H-NbSe$_2$, with root mean square surface roughness's around 4\rm{\AA} from AFM images, guarantee intimate contact to the GaAs substrate. All measurements in the temperature range from 5\thinspace K to 300\thinspace K were carried out using a custom low-noise shielded chamber mounted in a Quantum Design Physical Properties Measurement System (PPMS). Four-terminal in-plane transport measurements for intrinsic crystals were performed using lock-in techniques at 526\thinspace Hz. DC $I$-$V$ and AC $C$-$V$ characteristics at kilohertz frequencies were acquired by a Keithley 2400 source meter and HP 4284A LCR meter. More than 10 devices for both 1T-TiSe$_2$/\textit{n}-GaAs and 2H-NbSe$_2$/\textit{n}-GaAs junctions were tested and shown to exhibit reproducible features. In Figs.~\ref{supfigure1} (c) and (d), we plot the in-plane resistance as well as the derivative with respect to temperature for thin flakes of 1T-TiSe$_2$ and 2H-NbSe$_2$ as a function of temperature, the CDW transition temperatures $T_{\rm{CDW}}$ of 1T-TiSe$_2$ and 2H-NbSe$_2$ are respectively 193\thinspace K and 33\thinspace K.

\section{Schottky barriers as interface probes}

\begin{figure}
\includegraphics[width=0.4\textwidth]{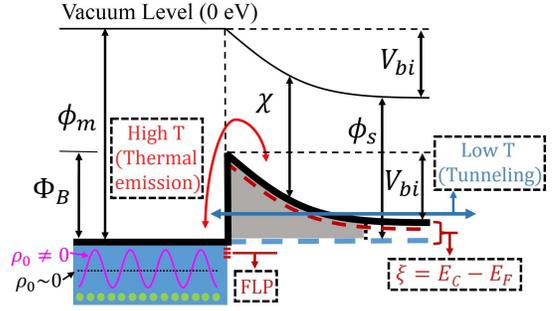}
\caption{Band diagram of a Schottky junction formed by a metallic layered material (with CDW phase ($\rho_0\not=0$, magenta) or without CDW phase ($\rho_0\sim0$, black-dashed)) in intimate contact with a Si-doped GaAs semiconductor where the donor impurity band (red dashed line) is close to the conduction band (solid black line). At high \textit{T} (low \textit{T}), thermionic emission (quantum tunneling) dominates. Barrier height: $\Phi_{B}$; Built-in potential: $V_{bi}$; Work function of layered materials: $\phi_m$; Electron affinity of GaAs: $\chi$; Work function  of GaAs: $\phi_{s}$; Fermi level pinning: FLP; CDW present(absent): $\rho_0 \neq 0(=0)$}
\label{figure1}
\end{figure}

As shown in Fig.\ref{figure1}, the measured built-in potential $V_{bi}$ ideally is equal to the difference between the work functions $\phi_m$ of the TMD metallic electrode and $\phi_s$ of the adjacent semiconductor\cite{thermionic}. In the presence of a correction for Fermi level pinning (FLP), the work function $\phi_s$ of the semiconductor is modified to be $\phi^{\prime}_{s}$ and $V_{bi}$ becomes,
\begin{equation}
V_{bi}=\phi_m-\phi^{\prime}_{s}=-\mu-\phi^{\prime}_{s} ,
\label{Vbi}
\end{equation}
where the chemical potential $\mu$ is referred to the vacuum level where $\mu = 0$. There are two contributions to a temperature-dependent $V_{bi}(T)$: the first derives from the semiconductor side where $\xi (T) = E_C - E_F = k_BT \ln (N_C/N_D)$ with $N_C (N_D)$ denoting the intrinsic (donor dopant) density and the second derives from the TMD side where the chemical potential $\mu$ or work function $\phi_m$ shifts associated with CDW formation occur.  

Accordingly, after normalizing out the well known $V_{bi}(T)$ changes arising from the temperature dependence of $\xi (T)$ in GaAs, we are able - as shown in the paragraphs below - to utilize Eq. \ref{Vbi} to isolate and study the chemical potential shifts due to CDW formation in our 1T-TiSe$_2$ electrodes. At low temperatures, 1T-TiSe$_2$ is like a heavily doped semiconductor with a small gap and can be treated as a metallic electrode for this analysis. Similar signatures are also observed in 2H-NbSe$_2$ based junctions indicating CDW-induced chemical potential shifts in 2H-NbSe$_2$. For 2H-NbSe$_2$, the  observed $V_{bi} (T)$ evolves with decreasing temperature, suggesting a crossover from short range to long range CDW order below 90\thinspace K in accord with transport and STM measurements\cite{nbse1}. Unexpectedly, for 1T-TiSe$_2$, the short-to-long range CDW formation feature also appears when we closely analyze the temperature-dependent chemical potential shift in accord with Eq.~\ref{Vbi}. This common feature for two different materials with demonstrably different CDW formation mechanisms is somewhat surprising and presents a unifying theme embracing the nature of second-order CDW phase transitions in layered 2D TMD materials.

\section{Results and discussions}
\begin{figure}
\includegraphics[width=0.48\textwidth]{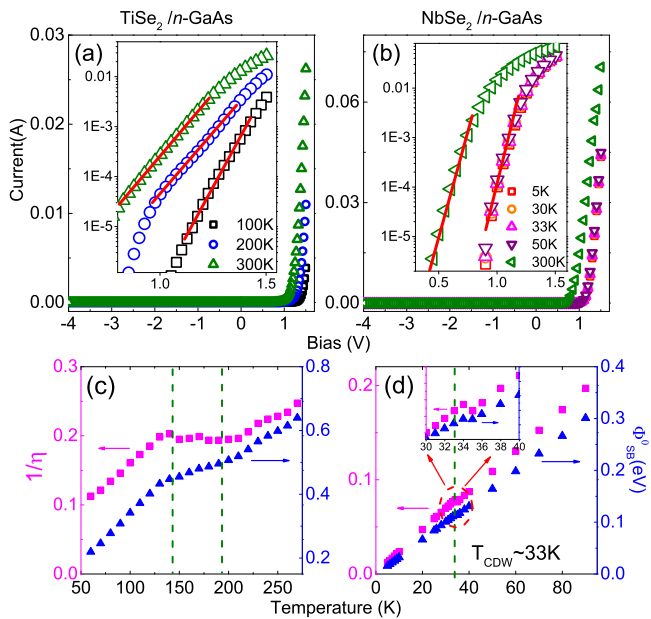}
\caption{(a)--(b): The $I-V$ characteristics of 1T-TiSe$_2$/\textit{n}-GaAs and 2H-NbSe$_2$/\textit{n}-GaAs junctions at selected temperatures. Insets: Semi-log scaled $I-V$ characteristics. Red solid lines indicate the theoretical fitting range. (c)--(d): Schottky barrier model fitted parameters $\Phi^{0}_{SB}$ and $\eta^{-1}$ as a function of temperature for 1T-TiSe$_2$/\textit{n}-GaAs and 2H-NbSe$_2$/\textit{n}-GaAs junctions. Green dashed lines delineate the temperature span of the kink signature of $\Phi^{0}_{SB}(T)$ and $\eta^{-1}(T)$. Inset of (d): Zoomed in plot in the temperature range of 30\thinspace K to 40\thinspace K.}
\label{figure2}
\end{figure}

According to transport measurement seen in Figs.~\ref{supfigure1}(c) and (d), the CDW transition temperatures $T_{\rm{CDW}}$ of 1T-TiSe$_2$ and 2H-NbSe$_2$ are respectively 193\thinspace K and 33\thinspace K, both of which are consistent with the literature on intrinsic bulk materials\cite{tise tCDW, nbse tCDW}. The $I$-$V$ characteristics for 1T-TiSe$_2$/\textit{n}-GaAs and 2H-NbSe$_2$/\textit{n}-GaAs junctions are respectively shown in Figs.~\ref{figure2}(a) and (b). The rectifying features 
indicate the formation of Schottky barriers when TMD-CDW materials are in intimate Van der Waals contact with GaAs substrates. The transport characteristics across Schottky barriers are well described by thermionic emission theory at high temperatures\cite{thermionic} where thermionic emission dominates. With phenomenological modifications\cite{Ang}, the thermionic emission equation is also capable of describing the transport characteristics of Schottky barriers at low temperatures where field emission tunneling is also significant as illustrated in the schematic of Fig.~\ref{figure1}. 

The temperature-dependent zero-bias barrier height $\Phi^{0}_{SB}(T)$ and ideality factor $\eta(T)$ characterizing the Schottky barrier profile can be extracted from the $I$-$V$ characteristics, described by $I\propto \rm{exp}\it{(-e\Phi^{0}_{SB}(T)/k_{B}T)}\cdot \rm{exp}\it{[(V-I\cdot R_s)/\eta(T)k_{B}T]}$\cite{werner}, where the contact resistance $R_s$ is negligible as it is for our junctions. These parameters $\Phi^{0}_{SB}(T)$ and $\eta^{-1}(T)$ for both 1T-TiSe$_2$/\textit{n}-GaAs and 2H-NbSe$_2$/\textit{n}-GaAs junctions, are extracted by fitting from the linear regime of semi-log scaled $I$-$V$ curves indicated in insets of Figs.~\ref{figure2} (a) and (b) and are shown in Figs.~\ref{figure2} (c) and (d) respectively. 

The deflection or kink delineated by slope changes seen in Fig.~\ref{figure2}(c) for the 1T-TiSe$_2$/GaAs junction in the 140\thinspace K and 200\thinspace K range reveals CDW-induced changes in $\Phi^{0}_{SB}(T)$ and $\eta^{-1}(T)$ corresponding to changes of the barrier profile. Such changes are the result of the transition from semimetal to semiconductor in 1T-TiSe$_2$ related to the CDW formation\cite{tise2}. A similar, although less pronounced, kink signature is also apparent in the vicinity of $T_{\rm{CDW}} \approx$ 33\thinspace K for the 2H-NbSe$_2$ as shown in Fig.~\ref{figure2}(d), thereby corroborating the correlation between barrier profile change and CDW formation in 2H-NbSe$_2$. 

\begin{figure}
\includegraphics[width=0.48\textwidth]{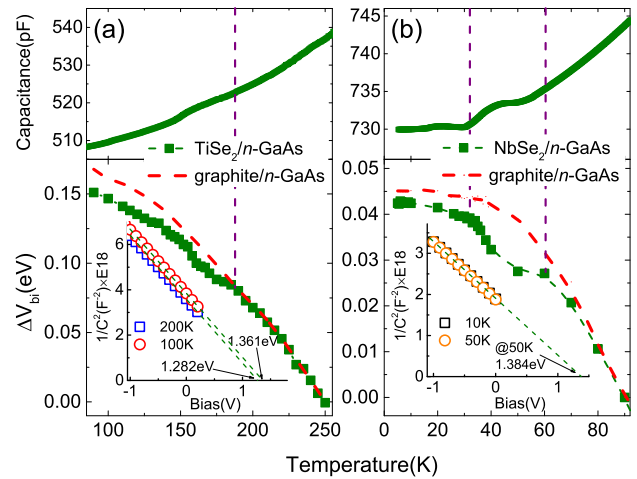}
\caption{Upper panel: Temperature dependent zero-bias capacitance; Lower panel: Temperature dependent relative built-in potential $\Delta V_{bi}$ for (a) 1T-TiSe$_2$/\textit{n}-GaAs and (b) 2H-NbSe$_2$/\textit{n}-GaAs junctions. The red dashed lines indicate the scaled $\Delta V_{bi}$ of a graphite/\textit{n}-GaAs junction. The reference temperatures are 250\thinspace K and 90\thinspace K for (a) 1T-TiSe$_2$/\textit{n}-GaAs and (b) 2H-NbSe$_2$/\textit{n}-GaAs junctions respectively. Insets of (a) and (b): $C-V$ characteristics at 10\thinspace kHz for both junctions at selected temperatures. Purple dashed lines are guidelines for eyes. Note: Error bars are not shown.}
\label{figure3}
\end{figure}

Complementary to forward-bias transport through the interface, capacitance measurements in reverse bias provide a \textit{direct measurement} of $V_{bi}$ and therefore a quasistatic method to characterize the depletion region and hence the Schottky barrier profile. Firstly, we notice the temperature-dependent zero-bias capacitance, as shown in the upper panels of Fig.~\ref{figure3}(a) and (b), experiences a step-like anomaly spanning the range 150\thinspace K-200\thinspace K for 1T-TiSe$_2$/\textit{n}-GaAs and begins at 33\thinspace K-60\thinspace K for 2H-NbSe$_2$/\textit{n}-GaAs junctions with an upper bound determined below with further analysis. 
Here, $V_{bi}$ at each temperature is extracted from extrapolated abscissa intercepts of linear $1/C^2$ versus reverse bias voltage $V$ plots\cite{thermionic} ($1/C^2\propto (V-V_{bi})$) shown for both 1T-TiSe$_2$/\textit{n}-GaAs and 2H-NbSe$_2$/\textit{n}-GaAs junctions at selected temperatures in the insets of Figs.~\ref{figure3}(a) and (b). 

Central to this paper is the understanding that measured built-in potentials $V_{bi}$ for both metal/semiconductor (see Fig.~\ref{figure1}) and heterojunctions reflect the difference between the work functions of the two materials in contact\cite{thermionic}. For our doped GaAs substrates we calculate that deep within the GaAs the energy difference between the conduction band minimum and the chemical potential (Fermi energy) $\xi (T)$ is small ($\xi \approx$ 60\thinspace mV at 300\thinspace K and near zero at low temperature) compared to $V_{bi}$, thereby justifying the use of the measured $V_{bi}$ as a good estimate of the Schottky barrier height. The work functions of 1T-TiSe$_2$ and 2H-NbSe$_2$ at room temperature are around 5.4\thinspace eV and 5.6\thinspace eV\cite{sci adv}, which for the 4.07\thinspace eV electron affinity of GaAs together with the Schottky-Mott law for ideal Schottky barriers\cite{mott} gives Schottky barrier heights ($\phi_{\rm{Mott}}$) of 1.33\thinspace eV and 1.53\thinspace eV for 1T-TiSe$_2$/\textit{n}-GaAs and 2H-NbSe$_2$/\textit{n}-GaAs junctions respectively. However, the barrier heights from our $V_{bi}$ measurements at room temperature are somewhat lower at 1.182$\pm$0.025\thinspace eV and 1.054$\pm$0.010\thinspace eV respectively (the term $\xi$ is also included), indicating the presence of Fermi level pinning (FLP) effects\cite{tung}.

To minimize these unavoidable FLP effects in our analysis, a certain temperature range with less FLP origins but overlapping the CDW transition regime is needed. 
The pinning coefficient $\gamma$ is calculated by using the FLP relation $\phi=\gamma \phi_{\rm{Mott}}+(1-\gamma) E_g/2$ with the charge neutrality level $\phi_{\rm{CNL}}\sim E_{g}/2$ for simplicity\cite{tung}, where $\phi$ is the barrier height from measurement, $\phi_{\rm{Mott}}$ is the barrier height from the Schottky-Mott law\cite{mott}, and $E_g = 1.42$\thinspace eV is the semiconductor band gap. As seen in the lower panels of Fig.~\ref{figure3}, the temperature-dependent relative built-in potentials $\Delta V_{bi}=V_{bi}(T)-V_{bi}(T_{ref})$ for both junctions are depicted as solid green squares. We note, as discussed earlier, that the background increase of $V_{bi}$ with decreasing $T$ is due to the monotonic changes in $\xi (T)$ in the semiconductor whereas the kink is due to CDW induced rearrangement of electron energy distributions in the TMD electrode.  For the 1T-TiSe$_2$/\textit{n}-GaAs junction, we choose the reference temperature $T_{ref}$ as 250\thinspace K, $V_{bi}(250\rm{K})$=1.208$\pm$0.005\thinspace eV and $\gamma\sim0.88$, whereas $T_{ref}=90$\thinspace K, $V_{bi}(90\rm{K})$=1.353$\pm0.003$\thinspace eV and $\gamma\sim0.79$ for the 2H-NbSe$_2$/\textit{n}-GaAs junction. As a procedural check of these similarly calculated pinning coefficients, the results here for the tested devices are not sensitive to the the chosen reference temperatures within the vicinity of $\pm$20\thinspace K.
  
To further interpret our observed step-like anomalies in $V_{bi}(T)$, we use graphite/\textit{n}-GaAs junctions as reference junctions that are characterized by the same method as already discussed. The red dashed lines shown in Fig.\thinspace\ref{figure3} are data scaled by manually shifting the measured $V_{bi}$ for the graphite/\textit{n}-GaAs junction, with $V_{bi}(250\rm{K})$=1.082$\pm0.008$\thinspace eV and $V_{bi}(90\rm{K})$=1.252$\pm0.005$\thinspace eV, to the scale of 1T-TiSe$_2$/\textit{n}-GaAs and 2H-NbSe$_2$/\textit{n}-GaAs junctions respectively. Here, we consider highly ordered pyrolytic graphite (HOPG) to serve as the ``ideal'' Van der Waals featureless electrode for the following two reasons: Firstly, HOPG is also a 2D-layered material with a similar honeycomb crystal structure of TMD materials and secondly, HOPG exhibits no phase transition within the temperature range of 5\thinspace K to 300\thinspace K.

Since metal-induced gap states (MIGS) are strongly suppressed in Van der Waals junctions\cite{sci adv}, we justifiably exclude the MIGS induced FLP effect at lower temperatures. In addition, since the lattice distortion's amplitude during CDW formation in 1T-TiSe$_2$ and 2H-NbSe$_2$ is small compared to the in-plane distance between Se/Ti(Nb)--Se atoms\cite{bond tise, bond nbse}, we can also rule out the change of barrier height due to significant breaking of chemical bonds at the Schottky barrier interface. Moreover, with carrier density much higher than moderately doped semiconductors for both materials\cite{nbse tCDW, tise GL}, we exclude the possibility of the observed changes originating from the modulation of interfacial dielectric properties during the CDW formation, a more pronounced mechanism that has already been reported in 1T-TaS$_2$ harboring first-order CDW phase transitions\cite{xc}.

We consequently interpret the $\Delta V_{bi}$ kink feature existing in both junctions as a chemical potential shift accompanying CDW formation in 1T-TiSe$_2$ and 2H-NbSe$_2$. Although the explicit correlation between CDW formation and chemical potential has already been discussed and confirmed in 1T-TiSe$_2$ owing to the change of the associated order parameter's amplitude during CDW formation\cite{tise condensate, tise condensate theory}, a corresponding  shift in 2H-NbSe$_2$ has to our knowledge not been reported.

In both 1T-TiSe$_2$ and 2H-NbSe$_2$, the CDW wave vector $\textbf{q}$ is temperature independent\cite{tise tCDW, nbse1, nbse2}. Therefore, only the CDW energy gap $\Delta$ and the relative phase $\phi$ of the charge displacements characterize the CDW formation in both materials\cite{gruner}. From ARPES studies of 1T-TiSe$_2$ at low temperatures\cite{tise condensate} the change of the energy gap $\Delta$ is associated with CDW formation raising the chemical potential by around 40\thinspace meV. However, a slight deviation of the experimental result from an exciton condensate model at temperatures ranging from 150\thinspace K to 200\thinspace K is noticed\cite{tise condensate, tise condensate theory}, which when combining with the signature of short range coherent CDW\cite{tise short1, tise short2}, indicates that changes in phase coherence are also significant for CDW formation in 1T-TiSe$_2$. The case for CDW formation in 2H-NbSe$_2$ is conversely, somewhat different. During the CDW formation in 2H-NbSe$_2$, phase coherence is gradually developed and completed down to $T_{\rm{CDW}}$ (from short range to long range CDW), whereas the amplitude of the CDW order parameter is well-defined with a finite amplitude over the temperature range starting from higher temperature as shown in a systematic work by Chatterjee \textit{et al}\cite{nbse2}.

\begin{figure}
\includegraphics[width=0.5\textwidth]{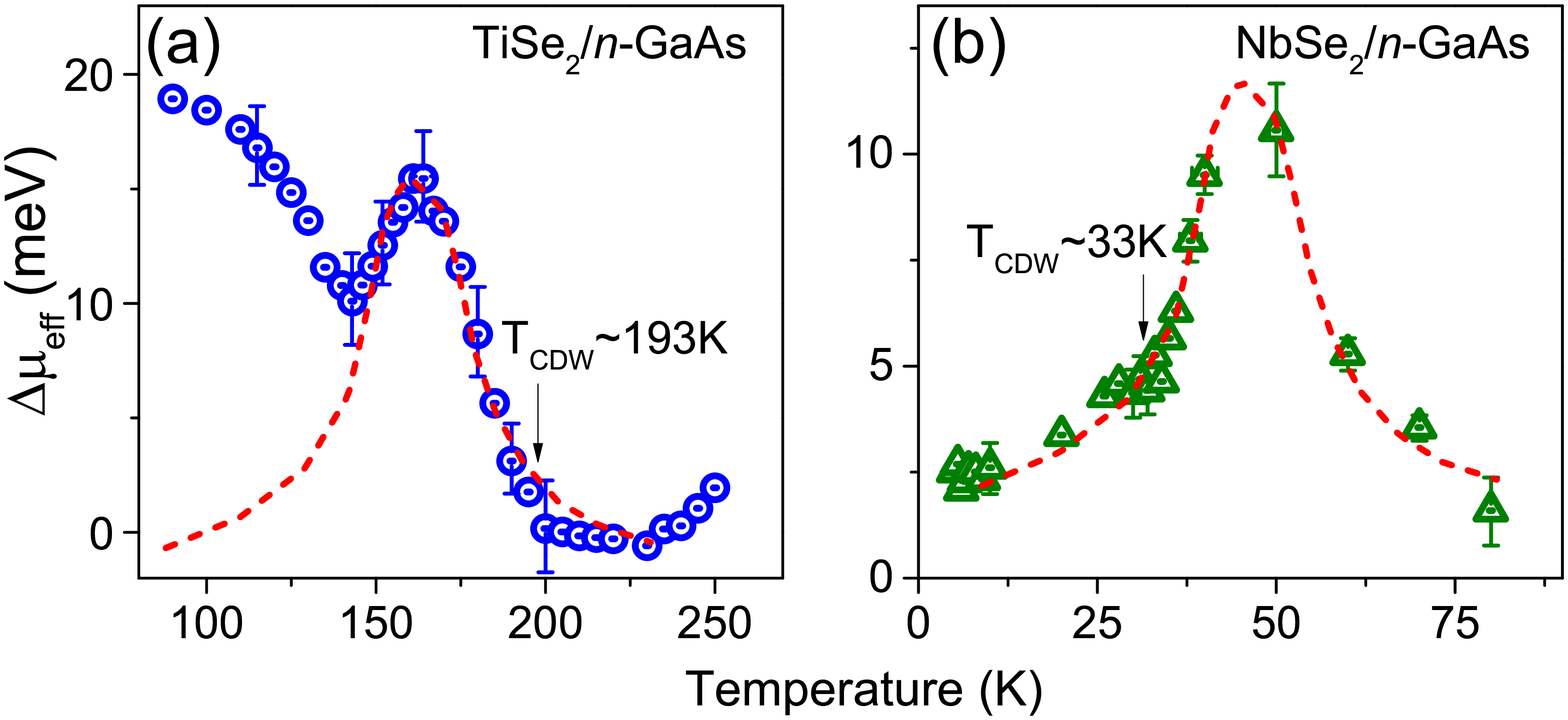}
\caption{Extracted temperature dependent effective chemical potential shift $\Delta \mu_{eff}$ of (a) 1T-TiSe$_2$ and (b) 2H-NbSe$_2$ from the measured built-in potential $V_{bi}$ of 1T-TiSe$_2$/\textit{n}-GaAs and 2H-NbSe$_2$/\textit{n}-GaAs junctions shown in Fig.\ref{figure3}. 
The red dashed lines are guides to the eye for the temperature dependent $\Delta \mu_{eff}$ proposed to be associated with the short (high temperature) to long (low temperature) range phase coherence of CDW formation.}
\label{figure4}
\end{figure}

In this paper we are addressing evidence in support of the hypothesis that chemical potential shifts in a CDW material are linked with the evolution of a CDW's complex order parameter comprising both amplitude and phase. To our knowledge there are no experiments that can simultaneously detect the amplitude and phase of the CDW order parameter. Our evidence (Fig.~\ref{figure4}) for effective chemical potential shifts $\Delta\mu_{eff}(T)$  is derived from the differences between the scaled $\Delta V_{bi}$ for graphite/\textit{n}-GaAs junction (red) and $\Delta V_{bi}$ for both junctions (green) shown in Fig.~\ref{figure3}, $\Delta\mu_{eff}(T)=V^{\rm{Gr}}_{bi}(T)-V^{\rm{CDW}}_{bi}(T)$. Inclusion of FLP effects does not change the monotonic relation between the change of Fermi level and barrier height. Accordingly, FLP effects clearly exist but do not modify our conclusions.

To gain a physical understanding of the peak features shown in Fig.~\ref{figure4}, we re-emphasize that CDW formation primarily affects $\phi_m$ and not $\phi^{\prime}_s$ of Eq.~\ref{Vbi}.  Hence, with decreasing temperature for both TMD materials, the high(low) temperature side of the respective peaks (red dashed lines) represents a simultaneous increase(decrease) in the quantities $V_{bi} (T)$, $\phi_m (T)$ and $-\Delta\mu_{eff} (T)$. The changes in $V_{bi}$ only manifest the changes in electronic energy associated with the redistribution of electrons in the Fermi sea and are not directly sensitive to Coulomb and strain energies. Accordingly, total energies are not measured. However, in the region of the peaks there is a delicate balance over a narrow temperature range where, on the low(high) temperature side of the peak, electronic energies are(are not) favored. For 1T-TiSe$_2$ photoemission measurements\cite{tise condensate} show a gradually increasing chemical potential upon cooling, in agreement with the overall trend in Fig.~\ref{figure4}(a) but without sufficient resolution to show the peak near 170\thinspace K. This \textit{increase} in chemical potential is interpreted in Ref.~\cite{tise condensate} as a temperature-dependent $\Delta$ during CDW formation mediated by excitons. A similar hump-like feature, peaked at around 50\thinspace K, is more pronounced in 2H-NbSe$_2$ as seen in Fig.\ref{figure4}(b) signifying common aspects of CDW formation in both 1T-TiSe$_2$ and 2H-NbSe$_2$. In both systems the phase coherence increases with decreasing temperature where homogeneous CDW domains with constant phase at low temperatures have evolved from locally isolated CDW domains dominating at higher temperatures\cite{nbse1}.

\section{Conclusions}

In summary, by use of ``simple'' Van der Waals Schottky barriers between layered TMD materials and the moderately doped semiconductor \textit{n}-type GaAs, we have found common features of CDW formation in 1T-TiSe$_2$ and 2H-NbSe$_2$. These common features suggest that similar considerations may apply to other layered TMD strong-coupling CDW materials. Our insights provide a complementary understanding of the mechanism of CDW formation in low-dimensional correlated systems and have shown a promising route for detecting and analyzing collective interfacial phenomena arising from strong correlation. The use of Schottky barriers as surface state probes not only enlightens our understanding of the electronic states in low dimensional correlated systems but also suggests technological applications.


\section{Acknowledgements}
The authors thank Andrew Rinzler for useful discussions and the Nanoscale  Research  Facility  at  the  University  of  Florida for access to cleanroom and rapid thermal annealing facilities. $^{\dagger}$A. J. Li and X. Zhu contributed equally to this work. This research work was supported by the National Science Foundation under Grant No. DMR–1305783 (AFH), the U.S. Army Research Office through the MURI Grant No. W911NF-11-10362 (LB) and the ONR-DURIP Grant No. 11997003 (LB).

\end{document}